\documentclass[]{iopart}

\usepackage{iopams}  
\usepackage{graphicx}
\usepackage[breaklinks=true,colorlinks=true,linkcolor=blue,urlcolor=blue,citecolor=blue]{hyperref}

\usepackage{booktabs}
\usepackage{dcolumn}

\begin{document}

\title[Long-range magnetic order in the 5$d^2$ double perovskite  Ba$_2$CaOsO$_6$]{Long-range magnetic order in the 5$d^2$ double perovskite  Ba$_2$CaOsO$_6$: Comparison with spin-disordered Ba$_2$YReO$_6$}

\author{C~M~Thompson$^{1,2}$, J~P~Carlo$^3$, R~Flacau$^4$, T~Aharen$^{1,5}$, I~A~Leahy$^3$, J~R~Pollichemi$^3$, T~J~S~Munsie$^6$,
T~Medina$^6$, G~M~Luke$^{6,2,7}$, J~Munevar$^8$, S~Cheung$^9$, T~Goko$^9$, Y~J~Uemura$^9$ and J~E~Greedan$^{1,2}$}
\address{$^1$Department of Chemistry and Chemical Biology, McMaster University, Hamilton, ON L8S 4M1, Canada}
\address{$^2$Brockhouse Institute for Materials Research, McMaster University, Hamilton, ON L8S 4M1 Canada}
\address{$^3$Department of Physics, Villanova University, Villanova, PA  19085  USA}
\address{$^4$Canadian Neutron Beam Centre, Chalk River Laboratories, Chalk River, ON K0J 1J0,  Canada}
\address{$^5$Department of Energy and Hydrocarbon Chemistry, Graduate School of Engineering, Kyoto University, Nishikyo-Ku, Kyoto 615-8510, Japan}
\address{$^6$Department of Physics and Astronomy, McMaster University, Hamilton, ON L8S 4M1,  Canada}
\address{$^7$Canadian Institute For Advanced Research, Toronto, ON, M5G 1Z8 Canada}
\address{$^8$Centro Brasiliero Pesquisas Fisicas (CBPF), Rio de Janeiro, Brazil}
\address{$^9$Department of Physics, Columbia University, New York, NY 10027 USA}
\ead{\mailto{jeremy.carlo@villanova.edu}}

\date{\today}

\begin{abstract}
The B-site ordered double perovskite Ba$_2$CaOsO$_6$ was studied by d.c. 
magnetic susceptibility, powder neutron diffraction and muon spin relaxation methods. 
The lattice parameter is $a$ = 8.3619(6) \AA\ at 280~K and cubic symmetry ($Fm\overline{3}m$) is 
retained to 3.5~K with $a$ = 8.3462(7) \AA.  Curie-Weiss susceptibility behaviour is 
observed for $T$ $>$ 100~K and the derived constants are C = 0.3361(3) emu-K/mole 
and $\Theta_{CW}$ = $-$156.2(3)~K, in excellent agreement with literature values. This Curie 
constant is much smaller than the spin-only value of 1.00 emu-K/mole for a 5$d^2$ 
Os$^{6+}$ configuration, indicating a major influence of spin-orbit coupling. Previous 
studies had detected both susceptibility and heat capacity anomalies near 50~K but no 
definitive conclusion was drawn concerning the nature of the ground state.  
While no ordered Os moment could be detected by powder neutron diffraction, muon 
spin relaxation ($\mu$SR) data show clear long-lived oscillations indicative of a continuous transition to
long-range magnetic order below $T_C$ = 50~K.  An estimate of the ordered moment on 
Os$^{6+}$ is $\sim$0.2 $\mu_B$, based upon a comparison with $\mu$SR data for Ba$_2$YRuO$_6$ 
with a known ordered moment of 2.2 $\mu_B$.   These results are compared with those 
for isostructural Ba$_2$YReO$_6$ which contains Re$^{5+}$, also 5$d^2$, and has a nearly 
identical unit cell constant, $a$ = 8.36278(2) \AA\ $-$ a structural doppelg\"anger. In contrast, Ba$_2$YReO$_6$ shows 
$\Theta_{CW}$ = $-$616~K, and a complex spin-disordered and, ultimately, spin-frozen ground 
state below 50~K, indicating a much higher level of geometric frustration than in 
Ba$_2$CaOsO$_6$.    The results on these $5d^2$ systems are compared to recent theory, which predicts
a variety of ferromagnetic and antiferromagnetic ground states.   In the case of Ba$_2$CaOsO$_6$,
our data indicate that a complex four-sublattice magnetic structure is likely.  
This is in contrast to the spin-disordered ground state in Ba$_2$YReO$_6$, despite a lack of
evidence for structural disorder, for which theory currently provides no clear explanation.      

\end{abstract}

\pacs{75.25.-j, 75.10.Jm, 75.40.Cx, 75.47.Lx, 75.50.-y, 75.70.Tj, 76.75.+i}

\maketitle

\section{Introduction}

Oxides with the double perovskite structure with rock-salt ordering of B cations 
\cite{anderson_dblperovskites} provide many opportunities for the 
design of new materials with the potential to exhibit geometric magnetic frustration. 
Such materials have the general composition A$_2$BB'O$_6$, where A is a large cation 
such as the divalent ions from Group 2 (Ca,Sr or Ba) or trivalent ions from Group 3 
(La and other large rare earth ions) and B and B' typically are smaller cations from the 
3$d$, 4$d$ and 5$d$ transition series in a variety of oxidation states and electronic 
configurations but can include smaller ions from Groups 2 and 3 such as Mg and Ca. 
Geometric magnetic frustration can arise when B is a diamagnetic ion and B' is magnetic
 as these sites are crystallographically distinct and each forms a face-centered 
cubic sublattice, which is equivalent to a network of edge-sharing tetrahedra as shown in 
Figure 1.  In the presence of antiferromagnetic nearest-neighbor correlations between B' cations
such a lattice exhibits frustration.

 \begin{figure}[h]
\includegraphics[width=90mm]{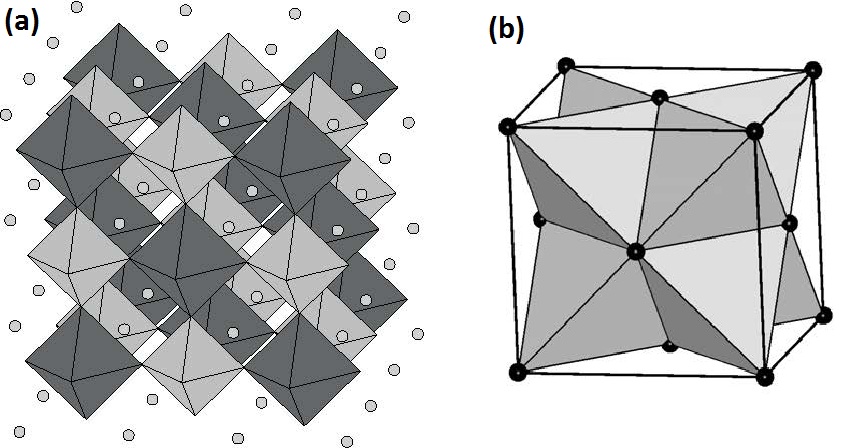}
 \caption{\label{}(a) The crystal structure of the B-site ordered double perovskite, A$_2$BB'O$_6$. 
The grey spheres, light grey octahedra and dark grey octahedra represent A ions, 
BO$_6$ octahedra and B'O$_6$ octahedra, respectively. (b) The geometrically frustrated face-
centered-cubic lattice of edge-sharing tetrahedra formed by the B' sites.}
 \end{figure}

The symmetry of the B' sublattice and the local point symmetry at the B' ion site are determined by 
Goldschmidt's tolerance factor, 
\begin{equation}
t = (\mathrm{A}-\mathrm{O}) / \sqrt{2}(<\mathrm{B,B'}>- \mathrm{O})
\end{equation}
 where A$-$O and 
$<$B,B'$>$$-$O  are the relevant cation $-$ oxide ion distances \cite{goldschmidt}. As $t$ decreases 
from the ideal value of 1, space group symmetries $Fm\overline{3}m$, $I4/m$ and $P2_1/n$ are observed sequentially, with 
respective B' site point symmetries of $m\overline{3}m$, $4/m$ and $\overline{1}$ \cite{howard}. The point symmetries determine 
the crystal field splitting of the t$_{2g}^n$ ground states. Of course the B' site electronic configuration 
confers a nominal spin, $S$, which can be strongly modified by the magnitude of the single ion spin orbit 
coupling (SOC) constant $\lambda$.  All of these factors $- $ sublattice symmetry, point symmetry, 
$S$ and SOC -- are “degrees of freedom” which can play a role in the determination of the magnetic ground 
state. Recent theory has attempted to take these factors into account at the 
mean field level \cite{balents_new,balents_s_1,dodds}.

In Table I some results for known double perovskites are catalogued. Some trends are evident. For $S$ $>$ 1, 
antiferromagnetic long range order (AFLRO) is almost always found. The frustration 
indices, defined as $f$ = $|\Theta_{CW}|$/$T_{\mathrm{order}}$, vary widely from 3 to 16. In some cases, for example 
Ba$_2$YRuO$_6$, there are anomalies. This material shows two susceptibility maxima 
and an unexpected gap of $\sim$5 meV in the inelastic neutron spectrum at $|Q|$ = 0.76 \AA$^{-1}$, 
the location of the (100) magnetic reflection. The presence of the gap has been attributed 
to the influence of SOC \cite{tomoko_3_2,carlo_BYRO}.  It is not known whether a similar feature exists 
for monoclinic La$_2$LiRuO$_6$.  For $S$ = 1, AFLRO seems rare, being known only when B'
 is the 3$d$ ion, Ni$^{2+}$ \cite{iwanaga}. The two cases with nominal $S$ = 1 for the 5$d$ ion, Re$^{5+}$, both show 
quite complex magnetic behaviour which has in the case of Ba$_2$YReO$_6$ been characterized 
by two distinct spin relaxation processes which lead ultimately to a spin-frozen ground state \cite{tomoko_1}.  
La$_2$LiReO$_6$ seems to feature a singlet ground state which remains poorly 
understood \cite{tomoko_1}.  For systems with nominal $S$ = $^1$/$_2$ a great variety of behaviours and 
ground states are observed including both ferromagnetic and antiferromagnetic long range order 
with minimal frustration for two materials involving Os$^{7+}$, Ba$_2$NaOsO$_6$ and Ba$_2$LiOsO$_6$ 
\cite{stitzer,erickson,steele}.  In contrast, Ba$_2$YMoO$_6$ exhibits the same $Fm\overline{3}m$ symmetry 
but its ground state is a unique, gapped singlet with $f$ $>$ 100 \cite{deVries_BYMO,tomoko_1_2,carlo_BYMO}.  
Upon lowering the symmetry to $P2_1/n$ in La$_2$LiMoO$_6$, both the high frustration and the 
singlet state disappear and a more ordered ground state emerges \cite{tomoko_1_2}.

\begin{table}[ht]
  \caption{Comparison of frustrated double perovskite systems. 
  $^*$This is the nominal spin quantum number assuming 
 L$-$S coupling. For 4$d$ and 5$d$ ions with $S$ = 1 and $^1$/$_2$ spin-orbit coupling will add a significant 
 orbital contribution which cannot be ignored.  
$^{**}$This is the frustration index, $f$ = $|\Theta_{CW}|$/$T_{cf}$. $T_{cf}$ refers 
 to the temperature below which either long range order, $T_c$, or apparent spin freezing, $T_f$, occurs.  B' symm.
refers to the point group symmetry at the magnetic B' site.
 AFLRO and FLRO refer to antiferromagnetic and ferromagnetic long range order. INC refers to long range
incommensurate magnetic order, and SRINC to short range incommensurate order.   
AFSRO signals short range antiferromagnetic order. SF  (spin frozen) indicates a non-LRO ground state 
in which static spins have been detected. SING? indicates a singlet ground 
 state as yet not well characterized. GSING refers to a gapped singlet ground state. 
$\lambda$ is the estimated 
 single-ion SOC constant for the magnetic ion in question.}    
  \begin{tabular}{rcrrrccrc}
   \br
      $S^*$ & Compound & $\Theta_{CW}$ (K) & $T_{cf}$ (K) &     $f^{**}$  & B' symm. & Ground State & $\lambda$ (meV) & Reference\\
   \mr
     5/2 & Ba$_2$MnWO$_6$ & $-$64     & 7.5     & 9      & $m\overline{3}m$  & AFLRO      &     $-$ & \cite{khattak,azad} \\
      5/2 & Sr$_2$MnWO$_6$  & $-$71     & 14      & 5      & $\overline{1}$      & AFLRO      &     $-$ & \cite{azad,munoz} \\
      3/2 & Ba$_2$YRuO$_6$   & $-$571   & 36      & 16     & $m\overline{3}m$ & AFLRO     & 200 & \cite{tomoko_3_2} \\
      3/2 & La$_2$LiRuO$_6$   & $-$170   & 23.8   & 7      & $\overline{1}$      & AFLRO     & 200 & \cite{tomoko_3_2} \\
      3/2 & La$_2$NaRuO$_6$ & $-$57     & 15       & 4      & $\overline{1}$       & INC          & 200 & \cite{aczel_LaNaBO, aczel_new} \\
      3/2 & La$_2$NaOsO$_6$ & $-$74     & 12       & 6      & $\overline{1}$       & SRINC          & 630 & \cite{aczel_LaNaBO, aczel_new}\\ 
     1     & Sr$_2$NiWO$_6$    & $-$175   & 54       & 3      &   $4/m$   & AFLRO    & 40 & \cite{iwanaga} \\
    1     & Ba$_2$YReO$_6$   & $-$616   & 35       & 18    &   $m\overline{3}m$ &   SF         & 590 & \cite{tomoko_1} \\
    1     & La$_2$LiReO$_6$   & $-$204   &     ?     &     ?   & $\overline{1}$        &  SING?    & 590 & \cite{tomoko_1} \\
      1/2 & Ba$_2$LiOsO$_6$   & $-$40    & 8         & 5      & $m\overline{3}m$   & AFLRO    & 630 & \cite{stitzer} \\
      1/2 & Ba$_2$NaOsO$_6$ & $-$10    & 6.8      &         $\sim$1   & $m\overline{3}m$     &  FLRO & 630 & \cite{stitzer} \\
      1/2 & Ba$_2$YMoO$_6$   & $-$219  & $<$2  & $>$400       & $m\overline{3}m$    & GSING & 135 & \cite{tomoko_1_2} \\
     1/2 & La$_2$LiMoO$_6$   & $-$45    &     4?    & 11    & $\overline{1}$         & AFSRO  & 135 & \cite{tomoko_1_2}\\
      1/2 & Sr$_2$CaReO$_6$  & $-$443  & 14       & 32    & $\overline{1}$         &   SF        & 590 & \cite{wiebe_SCRO} \\
      1/2 & Sr$_2$MgReO$_6$  & $-$426  & 45       & 9      & $\overline{1}$         &   SF        & 590 & \cite{wiebe_SMRO}\\
   \br
   \end{tabular}

   \label{tab:addlabel}
 \end{table}

In order to investigate further the possible systematics among this family of double perovskites, the material 
Ba$_2$CaOsO$_6$, containing 5$d^2$ Os$^{6+}$, which is isoelectronic with
Re$^{5+}$, has been studied in detail. This compound has been reported to crystallize in $Fm\overline{3}m$, a symmetry
 which is retained to 17~K according to x-ray diffraction results \cite{yamamura}. There is a remarkable 
accord in unit cell constant ($a$ = 8.359(5) \AA) with that for Ba$_2$YReO$_6$ ($a$ = 8.36278(2) \AA), and the 
importance of SOC should be very similar, making this material a true “doppelg\"{a}nger” 
to Ba$_2$YReO$_6$. Heat capacity and susceptibility data indicate an anomaly near 50~K but no firm conclusion 
was drawn concerning the nature of the ground state \cite{yamamura}.  Ba$_2$YReO$_6$ shows 
similar maxima in susceptibility and heat capacity at 25~K and 50~K \cite{tomoko_1}.  In this study the results of magnetic susceptibility, neutron powder diffraction and muon spin relaxation experiments are described 
and conclusions drawn regarding the nature of the ground state in Ba$_2$CaOsO$_6$.  Both materials are discussed
 in the context of recent theory on double perovskites \cite{balents_new,balents_s_1,dodds}.

\section{Experimental}

\subsection{Sample Preparation}

Ba$_2$CaOsO$_6$ was prepared by a conventional solid state reaction. Stoichiometric amounts of BaO$_2$, CaO,
 and Os metal were ground together, pressed into pellets and heated in air for 
30 mins at 1000$^{\circ}$C, in a platinum crucible. Due to evaporation of OsO$_4$, additional amounts of Os (10\% mol)
 were added. The sample was reground and re-pressed into a pellet than heated for an 
additional 30 mins at 1000$^{\circ}$C in air. This process was repeated once more but finally heated for 24 hr at 
1000$^{\circ}$C and a phase-pure sample was obtained. 

\subsection{X-Ray Diffraction}

Room temperature powder X-ray diffraction data were obtained on a PANanlytical X’Pert Pro diffractometer 
with an X’Celerator detector. Cu-K$\alpha_1$ radiation ($\lambda$ = 1.54056~\AA) with \
2$\theta$ step interval of 0.0167$^\circ$ was used for data collection. Rietveld refinements were performed 
using the FullProf suite \cite{fullprof}.

\subsection{Neutron Diffraction}

Neutron powder diffraction data were collected at the Canadian Neutron Beam Centre, Chalk River Nuclear 
Laboratories, using the C2 diffractometer with neutron wavelengths of 1.3305 \AA\   and 
2.3719 \AA\  at temperatures of 280~K and 4~K. 

\subsection{Magnetic Susceptibility}

A Quantum Design MPMS SQUID Magnetometer was used to perform bulk magnetization measurements. 
The magnetic susceptibility measurements were collected from 2~K to 300~K with applied 
field of 1000 Oe (0.10 T). Isothermal magnetization measurements were collected at 2, 25, and 100~K, 
from $-$5 to 5 T. 

\subsection{Muon Spin Relaxation}

Muon spin relaxation ($\mu$SR) data were collected in time-differential (TD) mode using the LAMPF 
spectrometer at the M20 surface muon beamline at TRIUMF, Vancouver, Canada.   $\mu$SR measurements 
serve as a probe of local magnetism, employing the fact that muons possess a sizable magnetic dipole 
moment.   A beam of initially 100\% spin-polarized positive muons ($\mu^+$) is 
produced, and muons are implanted one at a time into the sample.   Each muon comes to rest, 
typically at a a crystallographic interstitial site or near an oxygen anion.
The muon then undergoes Larmor precession in the local field at the muon site, with a frequency
$f$ = $\gamma$B, with $\gamma$ = 135.5~MHz/T, until it decays (with a characteristic 
2.2 $\mu$s timescale).  The muon then emits a positron, preferentially along the instantaneous spin axis 
at the time of decay.  These emitted positrons are detected by a pair of detectors positioned around the sample 
(often in front of, F, and behind, B, the sample) and the decay asymmetry is calculated as
\begin{equation}
	Asy = (B-F)/(B+F)
\end{equation}

The time dependence of this decay asymmetry follows that of the muon spin polarization $G_z$(t), and 
from this the internal local field distribution may be deduced.   Measurements performed in zero applied 
field are sensitive to extremely small internal magnetic fields (on the order of 1 G), including those due 
to nuclear magnetic moments, as well as magnetic fields which arise from the slowing down and static 
ordering of electronic moments.   Measurements taken in a weak applied transverse field (wTF) are used 
to calibrate the initial and baseline decay asymmetries, to account for detector efficiencies and 
areas.

A continuous beam of spin-polarized muons was implanted into a $\sim$1~g powder sample loaded into a $^4$He 
cryostat, and two-counter muon spin decay asymmetries were collected from 0 to 10 $\mu$s at 
temperatures ranging from 2~K up to 130~K.   Measurements were taken in zero field (ZF) and weak 
transverse field (wTF) modes using the ultra-low background sample holder. 

\section{Results and Discussion}

\subsection{Crystal Structure}

Rietveld refinement of both powder x-ray and neutron diffraction data at 280~K confirmed the cubic 
$Fm\overline{3}m$ structure, in accordance with the previously reported x-ray diffraction data \cite{yamamura}. 
The results pertaining to lattice constants, positional and displacement parameters, and goodness of fit
 are shown in Table II.  Figure 2 shows the results of the refinement at 3.5~K. The results from the
 refinement indicate  that, to within the resolution of the data, the phase retains cubic symmetry at 3.5~K 
with  $a$ = 8.3462(7) \AA. Simulations of plausible $I4/m$ models indicate that a tetragonal distortion 
as small as $c$/$a$ = 1.004 would have been easily detected. This is consistent with previous reports that 
$Fm\overline{3}m$ symmetry is retained to 17~K from x-ray diffraction data \cite{yamamura}.   No evidence for B-site
(Ca/Os) disorder was found.   This is consistent with earlier reports of $<$$3\%$ B-site disorder in similar double perovskites (\textit{e.g.} 
\cite{tomoko_3_2, tomoko_1, tomoko_1_2}, and the large difference in valence and ionic radius between 
Ca$^{2+}$ and Os$^{6+}$ solidly places Ba$_2$CaOsO$_6$ in the "rock salt ordered" region of the phase diagram of
Anderson \textit{et al.} \cite{anderson_dblperovskites}.

 \begin{figure}[h]
 \includegraphics[width=90mm]{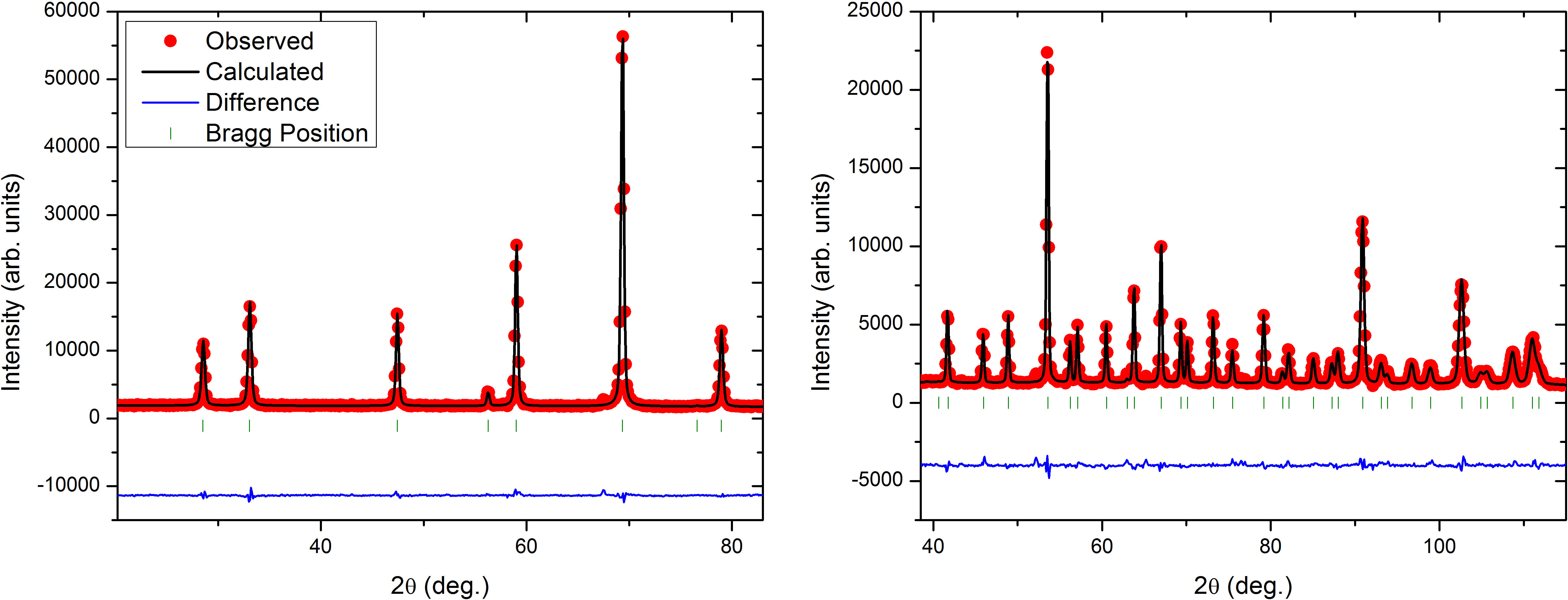}
 \caption{\label{}Low-temperature (3.5~K) powder neutron diffraction patterns of Ba$_2$CaOsO$_6$, measured 
with neutron wavelengths (a) $\lambda$ = 2.3719 \AA\ and (b) $\lambda$ = 1.3305 \AA.  }
 \end{figure}

 \begin{table}[ht]
  \indent
 \caption{The results from the Rietveld refinement of powder neutron data for Ba$_2$CaOsO$_6$ in 
 $Fm\overline{3}m$ at 280~K. The refinement was done simultaneously with two wavelengths from neutrons and one 
 from x-ray ($\lambda$ = 1.541 \AA). Included are selected interatomic distances for Ba$_2$CaOsO$_6$ at 280~K. Refinement results at 3.5~K, 
done simultaneously with two wavelengths from neutrons only, are reported in square brackets [ ] below
the 280~K data.}
\begin{tabular}{rrrrr}
    \br
          &       & 280~K &      &  \\
          &       & [3.5~K] &      &  \\ 
   \mr
          & x     & y     & z     & B$_{iso}$ (\AA$^2$) \\
    \mr
    Ba    & 0.25  & 0.25  & 0.25  & 0.587(52) \\
             &         &          &           & [0.024(37)]  \\
    Ca    & 0.5   & 0.5   & 0.5   & 0.704(105) \\
             &         &          &           & [0.193(85)]  \\
    Os    & 0     & 0     & 0     & 0.269(57) \\
             &         &          &           &  [0.25]    \\
    O     & 0.2294(23) & 0     & 0     & 0.966(50) \\
             & [0.22911(16)] &          &           & 0.412(33)]  \\
& & & & \\
    \textit{a} & 8.3619(6) &       &       &  \\
             & [8.3462(7)]  &          &           &                  \\
 \mr
 \mr
  & $\lambda$ =      &1.33 \AA & 2.37 \AA & 1.541 \AA   \\
 \mr 
  & $\chi^2$    & 2.91  & 1.98  & 2.78    \\
  &           & [4.16]         & [4.19]         &                             \\
  &  R$_p$    & 3.56  & 3.69  & 13.8   \\
  &           & [2.74]        & [2.74]         &                           \\
  &  R$_{wp}$   & 4.72  & 5.07  & 17.7    \\
  &           &  [3.84]       & [3.85]         &                            \\
  & R$_{exp}$  & 2.77  & 3.6   & 10.61   \\
  &           & [1.88]        & [1.88]         &                            \\
  &  R$_{Bragg}$ & 3.69  & 2.2   & 15.2    \\
  &           & [3.23]        & [1.20]         &                            \\
  &  R$_F$    & 2.26  & 1.68  & 12.3    \\
  &       &  [2.12]       & [0.76]         &                            \\
    \mr    
    \mr
&   & Ba$-$O  & 2.9613(2) &                \\
&     &        & [2.9561(8)]        &                                      \\
& Bond (\AA)&    Ca$-$O  & 2.2619(17) &                \\
&           &  & [2.2610(6)]        &                                       \\
&   & Os$-$O  & 1.9191(17) &                \\
&     &        & [1.9123(3)]        &                                      \\
    \br
    \end{tabular}
\end{table}

\begin{figure}[h]
 \includegraphics[width=85mm]{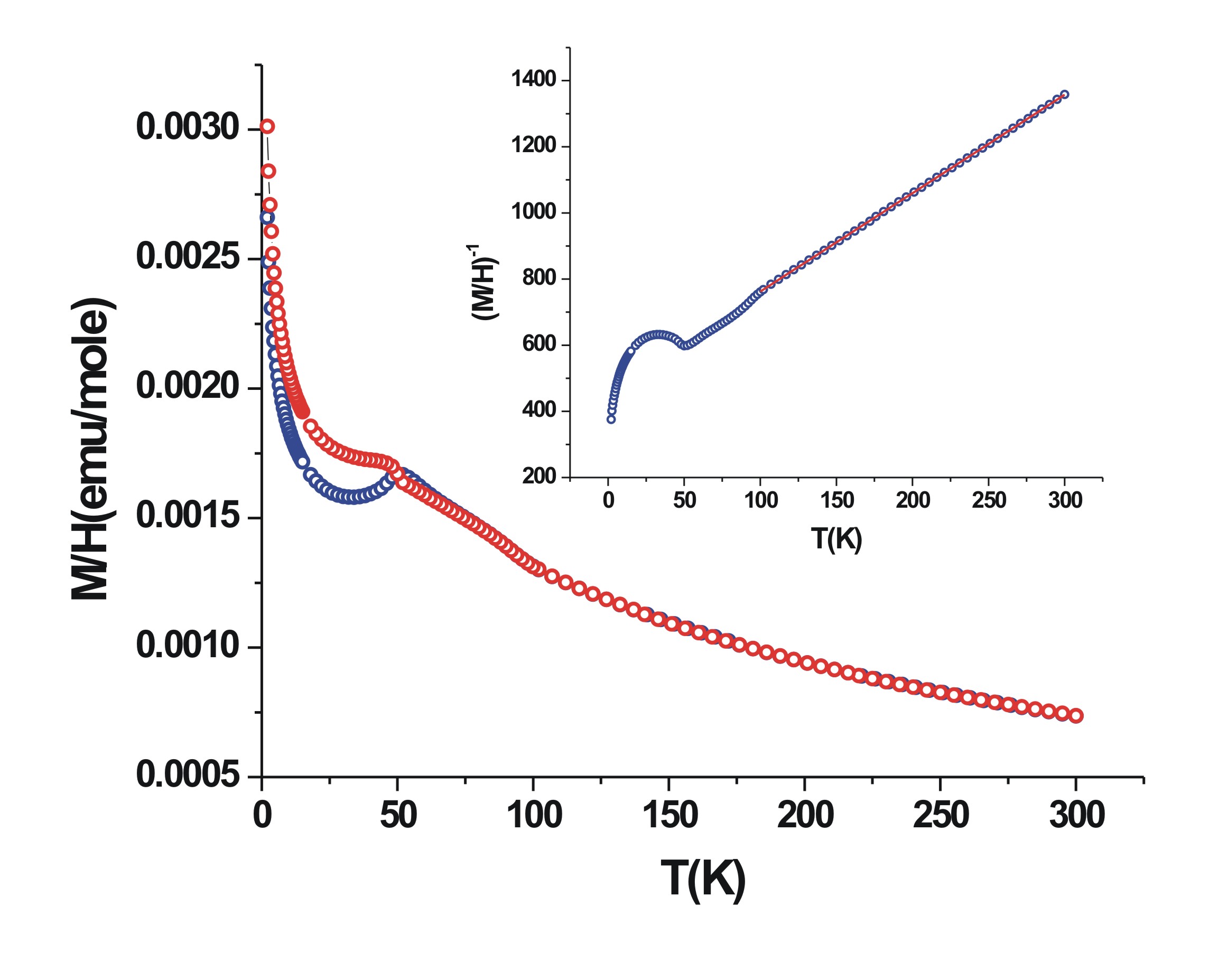}
 \caption{\label{} d.c. susceptibility data for Ba$_2$CaOsO$_6$.   Field cooled curve (FC, red) and
Zero-field cooled curve (ZFC, blue).   The inset shows the inverse susceptibility
 data fitted to the Curie-Weiss law (red line) yielding the constants given in the text.}
 \end{figure}

 \begin{figure}[h]
 \includegraphics[width=85mm]{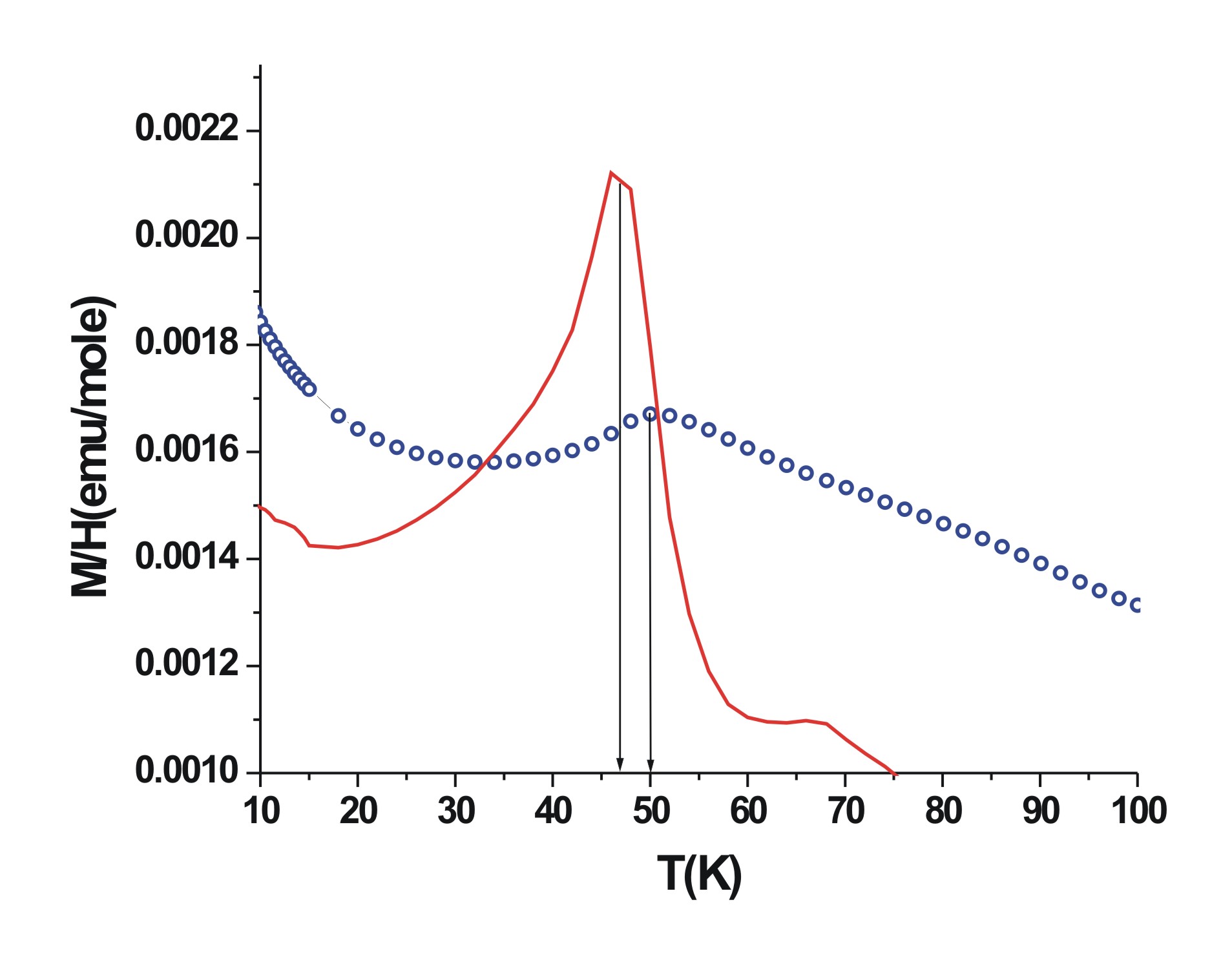}
 \caption{\label{} Low temperature ZFC data (blue) for Ba$_2$CaOsO$_6$.  Superimposed is a plot of the Fisher 
heat capacity d($\chi$$T$)/d$T$ vs. $T$ (solid red line), which can be expected to scale as the magnetic
contribution to the heat capacity \cite{fisher}, indicating a magnetic phase transition between 47 and 50~K.}
\end{figure}

\subsection{Magnetic Susceptibility}

In Figure 3 we show d.c. susceptibility data for the Ba$_2$CaOsO$_6$ sample studied here. The inset 
shows a fit to the Curie-Weiss law for $T$ $>$ 100~K which yields $C$ = 0.3361(3) emu-K/mole 
[$\mu_{\mathrm{eff}}$ = 1.640(1)~$\mu_\mathrm{B}$], which is much smaller than the spin-only value of 1.00 emu-K/mole  [2.83 $\mu_\mathrm{B}$], indicating a large influence for SOC, and $\Theta_{CW}$ = $-$156.2(3)~K. 
These results are nearly identical to those reported earlier \cite{yamamura}. 
 Figure 4 shows the low-temperature ZFC data upon which is superimposed Fisher's heat capacity
d($\chi$$T$)/d$T$ vs. $T$, which is expected to exhibit similar variation to the magnetic contribution to the
specific heat  \cite{fisher}, indicating a likely magnetic phase transition between 47~K and 50~K.

The field dependence of the magnetization is shown in Figure 5 at 100~K, 25~K and 2~K.  While linear behaviour with 
negligible hysteresis is seen for the two higher temperatures, the 2~K data indicate a small hysteresis 
and non-linearity. 

These results should be contrasted with those for Ba$_2$YReO$_6$, where the Curie-Weiss 
constants are $\mu_{\mathrm{eff}}$ = 2.105~$\mu_\mathrm{B}$ and $\Theta_{CW}$ = $-$616(7)~K \cite{tomoko_1}.  While 
$\mu_{\mathrm{eff}}$  is significantly closer to the spin-only value of 2.83 $\mu_\mathrm{B}$, indicating a 
smaller SOC influence, $\Theta_{CW}$ is four times larger, a remarkable
 result which indicates a much greater net antiferromagnetic exchange and, subsequently, a 
much more significant role for geometric frustration. The low-temperature susceptibility data for Ba$_2$YReO$_6$
 also show more complex behaviour with two rather broad maxima at 25~K and 50~K. Heat capacity
 data show two broad maxima at the same temperatures, which is not generally indicative of long-range order (more 
typically indicated by a $\lambda$-type anomaly); for example, the spin glasses Sr$_2$CaReO$_6$ \cite{wiebe_SCRO}
and Sr$_2$MgReO$_6$ \cite{wiebe_SMRO} both exhibit broad heat capacity anomalies.

 \begin{figure}[h]
 \includegraphics[width=85mm]{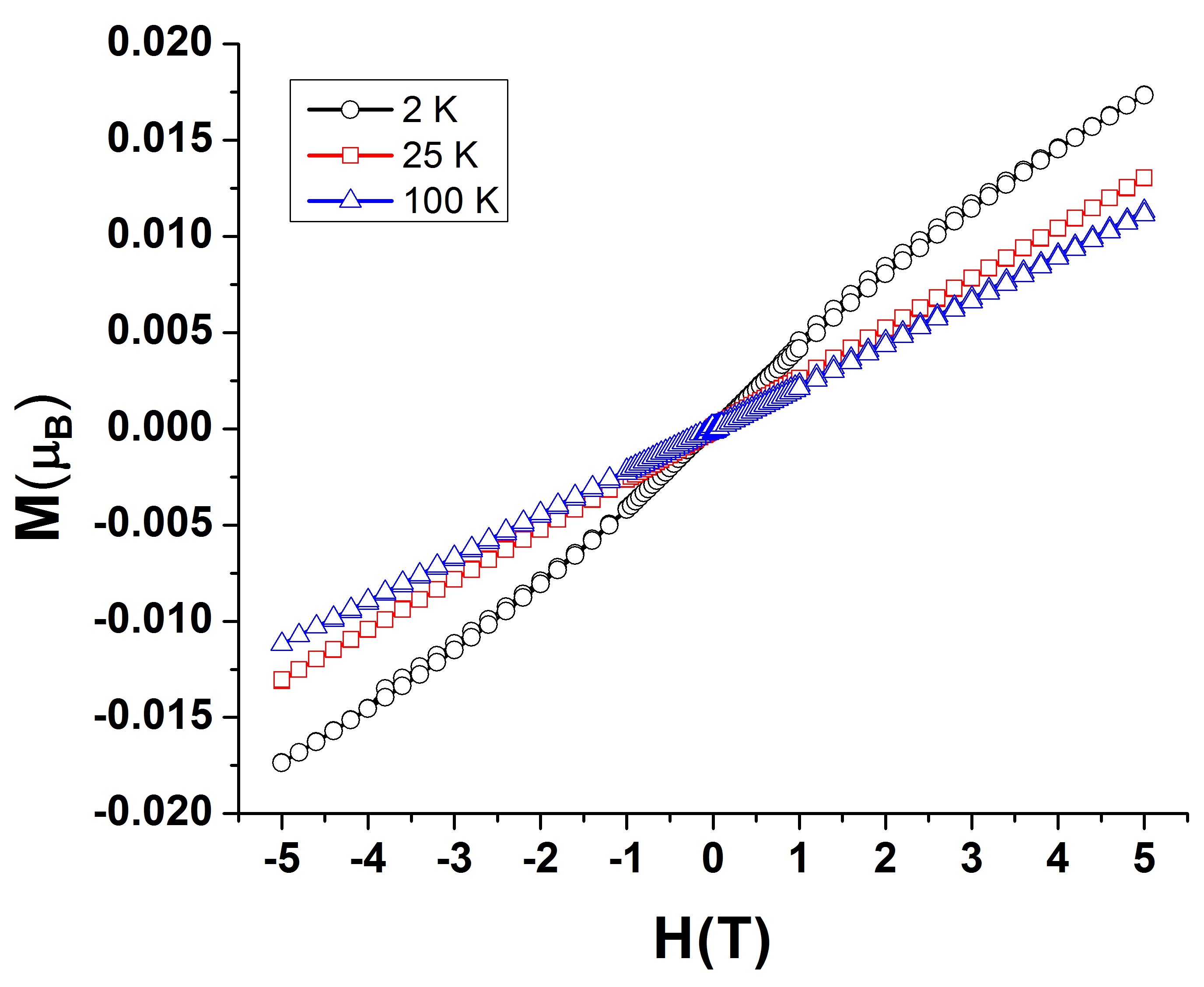}
 \caption{\label{}Hysteresis loops for Ba$_2$CaOsO$_6$ at 2, 25 and 100~K. }
 \end{figure}

\subsection{Magnetic Neutron Diffraction}

Figure 6 shows the 280~K and 4~K neutron diffraction patterns along with the difference intensity.
  Even with very long counting times, 63 hours at 3.5~K, no evidence for magnetic 
reflections was found between 10 and 85 degrees 2$\Theta$. As a small ordered moment is expected, simulations were
 carried out assuming a Type I f.c.c. magnetic structure as found for Ba$_2$YRuO$_6$ 
for a range of ordered moments on the Os site \cite{carlo_BYRO}.   For moments lower
 than 0.7 $\mu_\mathrm{B}$/Os the ratio (100)$_\mathrm{mag}$/(111)$_\mathrm{nucl}$ falls
 below 1\% and this is assigned as the likely upper limit of a detectable moment. In 
another approach, using the same instrument under the same experimental conditions, a 
moment of 0.3~$\mu_\mathrm{B}$/Ti$^{3+}$ ion was detected in the perovskite NdTiO$_3$ \cite{sefat}, 
 Taking into account the unit cell volume (243 \AA$^3$) and the number of 
magnetic ions per unit cell (4) a moment density of $\sim$5~$\times$~10$^{-3}$~$\mu_B$/\AA$^3$ is
 estimated as the upper limit. For Ba$_2$CaOsO$_6$ (cell volume 581 \AA$^3$) 
similar considerations also suggest a moment limit of $\sim$ 0.7$ \mu_\mathrm{B}$/Os ($\sim$5~$\times$~10$^{-3}$~
$\mu_B$/\AA$^3$).  Thus, it is possible to state with reasonable confidence that 
any ordered moment on Os$^{6+}$, assuming Type I f.c.c. long range order, must be less than 0.7 $\mu_\mathrm{B}$. 

 \begin{figure}[ht]
 \includegraphics[width=85mm]{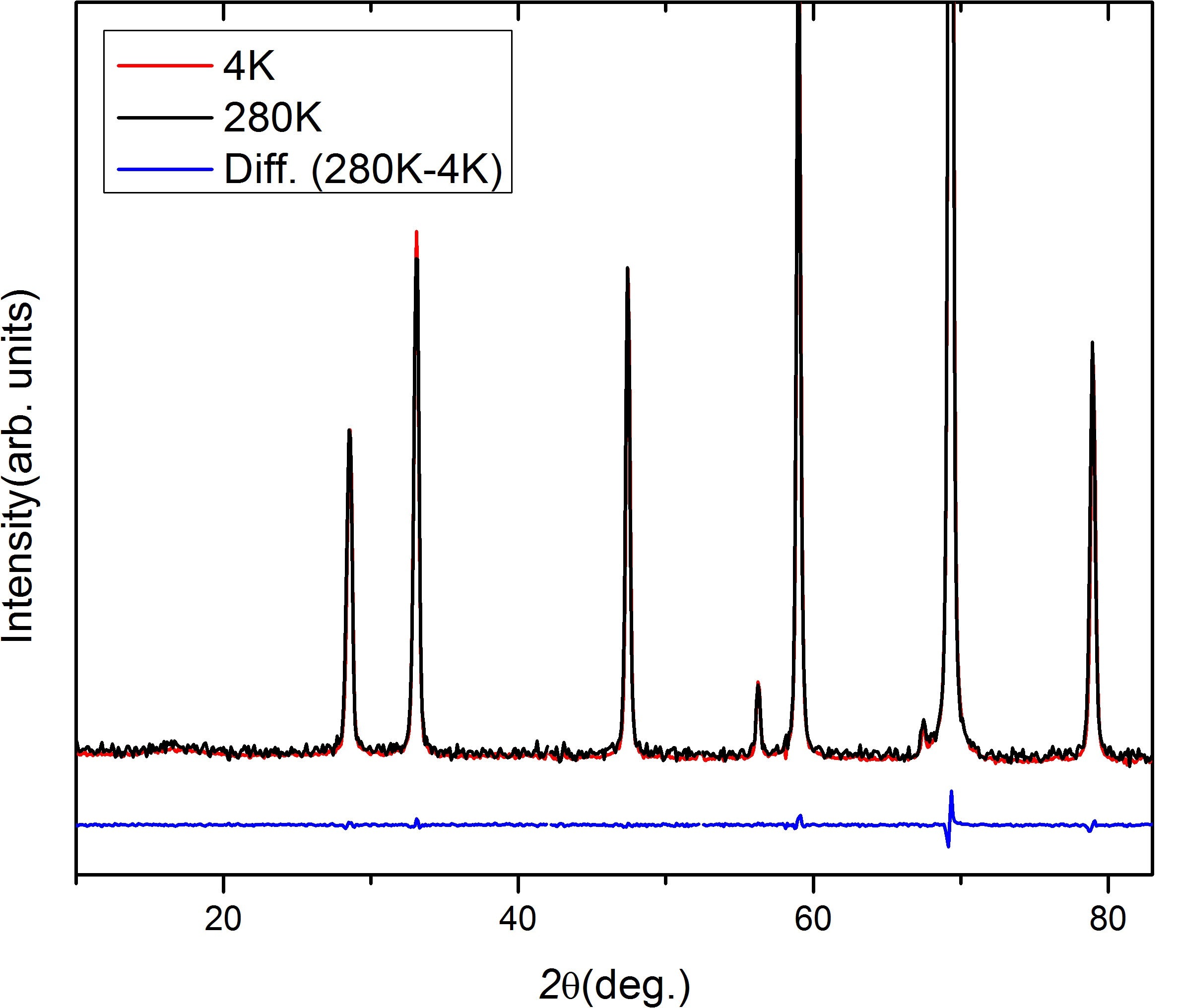}
 \caption{\label{}Neutron diffraction patterns at 4~K and 280~K, along with the difference between the two patterns.  
The counting time for the 4~K data was 63~hr.  Arrows indicate the locations of the [100] and [110] magnetic Bragg reflections,
as expected for Type I f.c.c. antiferromagnetic order; no intensity differences at these or any other locations are seen.}
 \end{figure}

\subsection{Muon Spin Relaxation}

Our ZF-$\mu$SR spectra very clearly show the onset of rapid relaxation and precession below 50~K, as seen in Figure 7.    
By base temperature, a clear and long-lived precessing asymmetry persists to at
 least 4~$\mu$s, with a frequency of 0.81~MHz.   

 \begin{figure}[ht]
 \includegraphics[width=85mm]{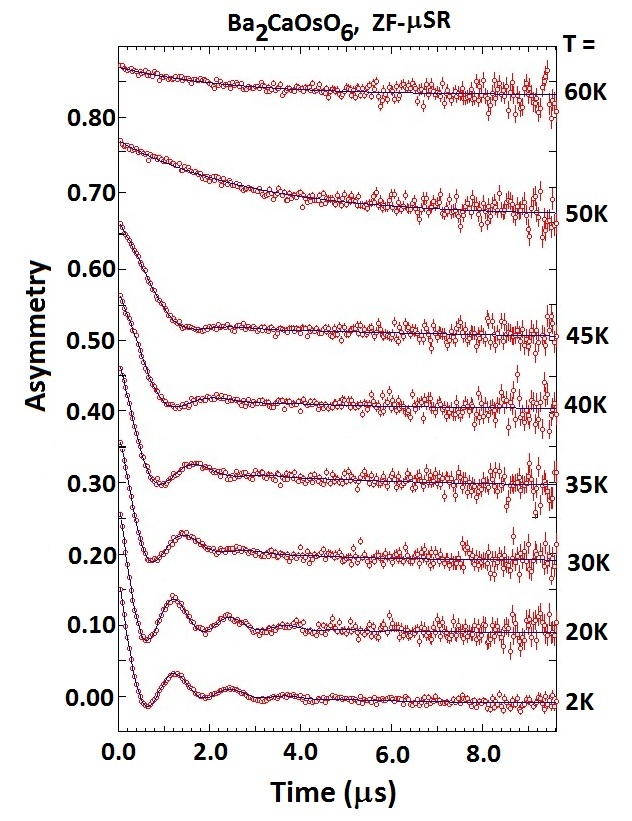}
 \caption{\label{}Zero-Field (ZF) muon spin relaxation data, fit to the function in Eq. 3.  Rapid relaxation and muon
 spin precession are visible below the ordering temperature 50~K; the existence of long-lived
 precession at low temperatures is indicative of long-range order.   The vertical axis depicts raw asymmetry, with 
offsets of 0.1 added to separate the traces.}
 \end{figure}

The entire muon spin precession signal is fit to a sum of precessing and relaxing asymmetries:

\begin{equation}
	Asy(t) = A_1e^{(-\lambda_1t)}\mathrm{cos}(2\pi ft) + A_2e^{(-\lambda_2t)} + A_3e^{(-\lambda_3t)} 
\end{equation}

In this fit function, the first two terms correspond to rapid precessing and non-precessing relaxing 
asymmetries present below the ordering temperature, whereas the third term corresponds to a 
slow relaxation seen at all temperatures.   Figure 8 shows the temperature dependence of the precession
 frequency $f(T)$, fit to a power-law temperature dependence 

\begin{equation}
f(T) = f_0(1-(T/T_C))^\beta 
\end{equation}

\noindent 
over the range 30~K $<$ T $<$ 50~K with an exponent $\beta$ fixed to the Heisenberg value 0.362, yielding 
$T_C$ = 48.4~K.  There are not enough data points to clearly distinguish between Ising, XY and Heisenberg
critical exponent values.   The three relaxation 
rates $\lambda_1$, $\lambda_2$ and $\lambda_3$ are depicted in Figure 9; due to interplay between 
the three terms, particularly near the ordering temperature, a “total relaxation” 
$\lambda_T$ = $\lambda_1$ + $\lambda_2$ + $\lambda_3$ is also plotted.   Near base temperature, the
 “fast” non-precessing relaxation rate $\lambda_2$ flattens out at approximately 
6.5 $\mu$s$^{-1}$, while the precessing relaxation rate $\lambda_1$ levels out at 1 $\mu$s$^{-1}$.   
The slow relaxation rate $\lambda_3$ remains at around 0.3 to 0.4 $\mu$s$^{-1}$ throughout the ordered region.
Note that these results (Fig. 8) are consistent with a continuous phase transition.

 \begin{figure}[ht]
 \includegraphics[width=85mm]{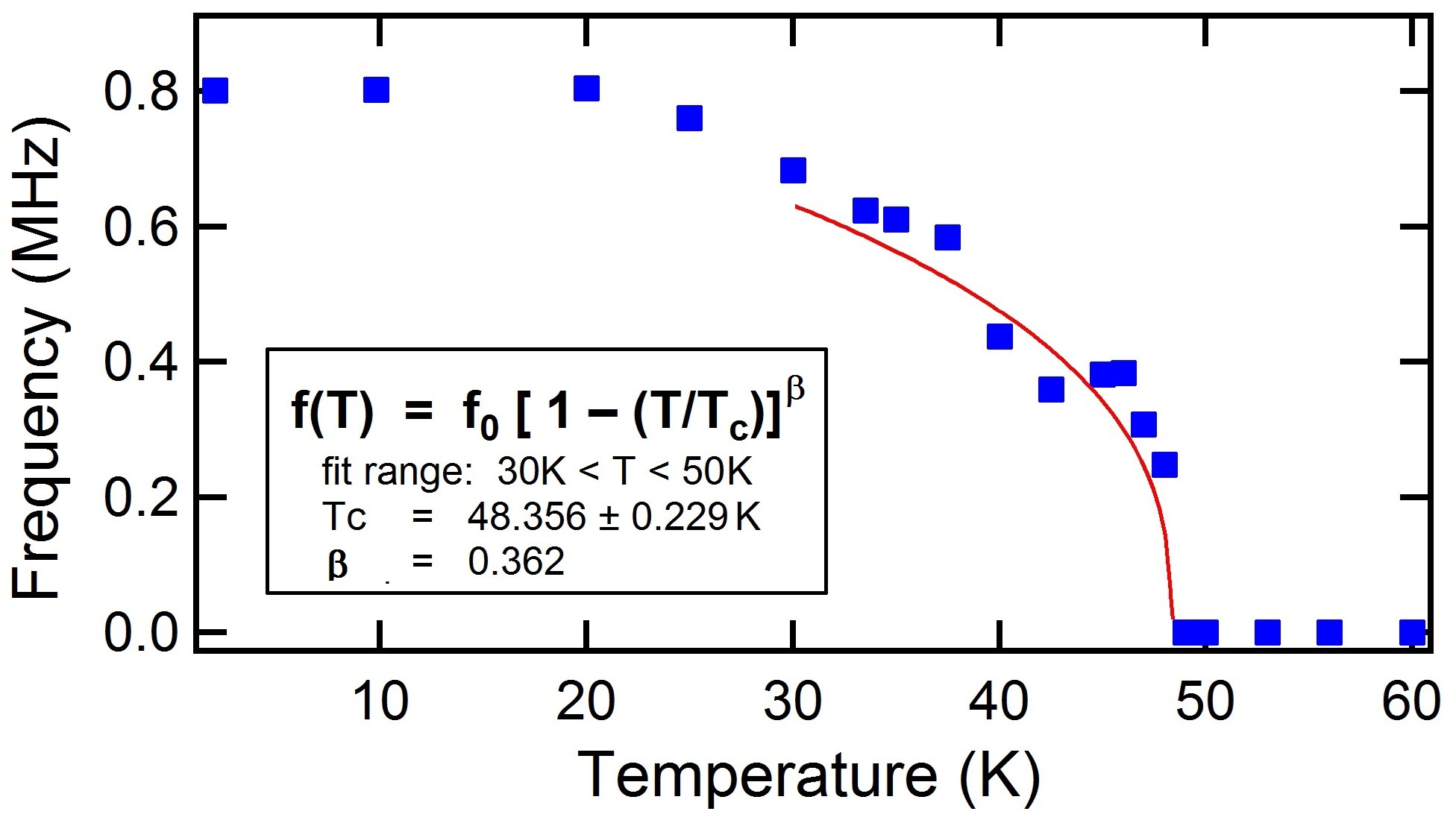}
 \caption{\label{}Muon spin precession frequency $f$ as a function of temperature.   The precessing signal 
abruptly becomes visible below the ordering temperature, and the frequency has been fit over the temperature
range 30~K $<$ T $<$ 50~K to a power-law 
dependence $f$($T$) = $f_0$(1 - $T$/$T_C$)$^\beta$), with $T_C$ = 48.4~K and $\beta$ fixed to the Heisenberg
value of 0.362.}
 \end{figure}

 \begin{figure}[ht]
 \includegraphics[width=85mm]{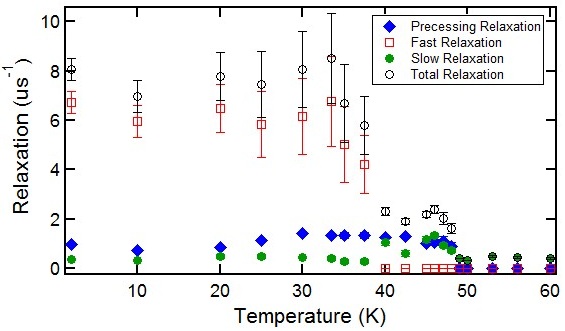}
 \caption{\label{}Muon spin relaxation rates $\lambda_1$ (“precessing”), $\lambda_2$ (“fast”) and
 $\lambda_3$ (“slow”) as described by the fit function in Eq. (2).   Due to interplay between the terms near the 
ordering temperature, a total relaxation $\lambda_T$, defined as the sum of the three individual 
relaxation rates, is also plotted.}
 \end{figure}

Such long-lived muon spin precession is associated with a highly homogeneous field distribution at the 
muon site, with $f$ = $\gamma$$B$, with the muon gyromagnetic ratio $\gamma$ = 135.5~MHz/T,
corresponding to an internal field at the muon site of 60 G. 
   
Muon spin relaxation data were collected under identical conditions for the antiferromagnetically long-range 
ordered isostructural material Ba$_2$YRuO$_6$ (Ru$^{5+}$  4$d^3$), known to exhibit an ordered 
moment size of 2.2 $\mu_\mathrm{B}$, as shown in Figure 10.   In this material, a similar relaxation function was used, 
although the precessing frequency reached 45.5~MHz at base temperature (with evidence for a second 
precessing frequency of 25-27~MHz), corresponding to an internal field at the muon site of 3.35~kG, a 
factor 57 times larger than in Ba$_2$CaOsO$_6$.   However, the relaxation rates $\lambda_1$ and $\lambda_2$ 
reach values of about 10 $\mu$s$^{-1}$ and 65.5 $\mu$s$^{-1}$ at T = 2~K, only a factor of 10 higher 
than in Ba$_2$CaOsO$_6$.    

 \begin{figure}[ht]
 \includegraphics[width=85mm]{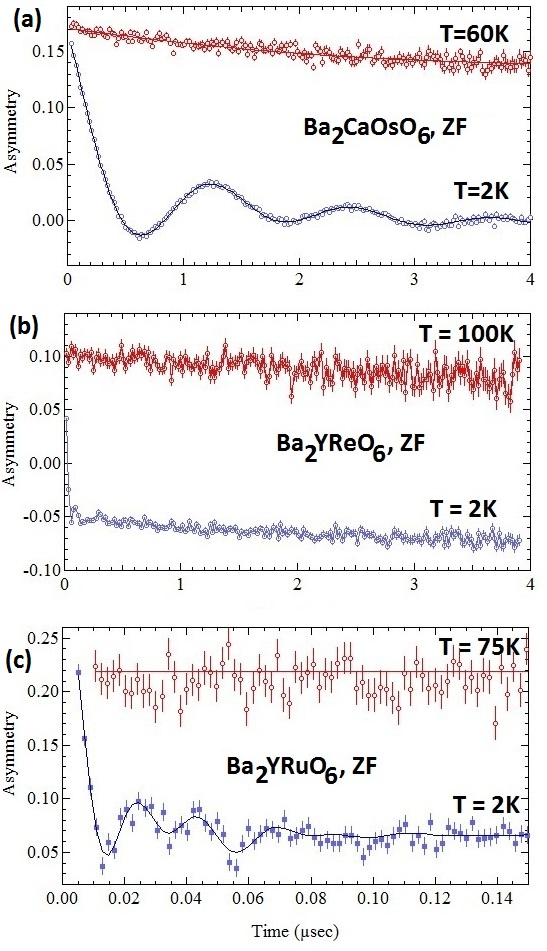}
 \caption{\label{}Comparison of ZF-$\mu$SR data for the isostructural systems Ba$_2$CaOsO$_6$ (a), 
Ba$_2$YReO$_6$ (b) and Ba$_2$YRuO$_6$ (c).   Note the much shorter timescale in the Ba$_2$YRuO$_6$ panel.   
Ba$_2$YRuO$_6$, known to exhibit commensurate long-range antiferromagnetic order, and 
Ba$_2$CaOsO$_6$ both exhibit clear precession of muon spins, corresponding to a relatively homogeneous and static
 internal field, indicative of long-range magnetic order.   Ba$_2$YReO$_6$, in contrast, exhibits rapid 
relaxation without sustained precession characteristic of a frozen but spatially disordered ground state.}
 \end{figure}

Previously, muon spin relaxation data have been reported on the isostructural and isoelectronic material
Ba$_2$YReO$_6$ (Re$^{5+}$ 5$d^2$), which is known to exhibit a glassy ground state \cite{tomoko_1},
 also shown in Figure 10.   In contrast to Ba$_2$YRuO$_6$ and Ba$_2$CaOsO$_6$, Ba$_2$YReO$_6$ 
exhibits only rapid relaxation with a single half-oscillation at low temperatures, as expected for a static glassy
 system \cite{Uemura_CuMn}, with no evidence for sustained precessing behaviour.   We therefore determine
 that Ba$_2$CaOsO$_6$, like Ba$_2$YRuO$_6$ and unlike Ba$_2$YReO$_6$, exhibits long-range order 
below its ordering temperature $T_C$ = 50~K which is of course consistent with susceptibility and heat capacity
data.

Further, since muon spin precession and relaxation rates scale linearly with the internal field, we determine
 that the internal field strengths in Ba$_2$CaOsO$_6$ are at least a factor of 10 times weaker than in 
Ba$_2$YRuO$_6$.   If the ordered moment distribution in the two is identical, we may determine an upper limit
 on the ordered moment size in Ba$_2$CaOsO$_6$ of 0.2 $\mu_\mathrm{B}$, based on the 2.2~$\mu_\mathrm{B}$ 
ordered moment in Ba$_2$YRuO$_6$ \cite{carlo_BYRO}.   However, $\mu$SR is a local probe of magnetism
 so the spatial range of the magnetic order in Ba$_2$CaOsO$_6$ cannot be determined independently by this method.   Indeed, 
muon spin precession has been seen in materials exhibiting short-range magnetic order \cite{wiebe_Li2Mn2O4},
 provided that the ordering correlation lengths are “long” relative to the spatial scale accessible 
to the muon.  Nonetheless, as has been emphasized, heat capacity and susceptibility data clearly establish 
long range order in this material.   Note that the estimated ordered moment size of 0.2~$\mu_\mathrm{B}$ per Os$^{6+}$ 
is consistent with similar moment sizes seen in other double perovskite antiferromagnets such as Ba$_2$NaOsO$_6$ \cite{erickson}.
For such small ordered moments, essentially undetectable to neutron powder diffraction, $\mu$SR provides nearly the 
only means of measurement.  In particular, magnetic Bragg peaks are expected to scale quadratically with the ordered moment size; 
an ordered moment of 0.2~$\mu_\mathrm{B}$ would yield a Bragg peak 12 times smaller than the 0.7~$\mu_\mathrm{B}$ upper
limit derived from 63~hr of neutron data.

It is also intriguing that both Ba$_2$YRuO$_6$ and Ba$_2$YReO$_6$ exhibit significantly higher frustration indices
$f$ than Ba$_2$CaOsO$_6$, indicating an increased level of frustration in the former materials.
 This is particularly intriguing in the case of the rhenate since in addition to being identical in structure, Re$^{5+}$ 
and Os$^{6+}$ are isoelectronic and, being neighbors on the periodic table, should have a very similar 
spin-orbit coupling parameter $\lambda$, which scales as $Z^4$. 

\subsection{Comparison with Theory}

Recently, a mean field theory of ordered cubic double perovskites based on $d^2$ ions under strong SOC has 
been published \cite{balents_s_1}.  A rather complex phase diagram is found with seven potential ground states which is 
reproduced as Figure 11.  The y-axis depicts J'/J, where J' is a nearest-neighbor FM interaction scaled to a J = 1 
AFM nearest-neighbor interaction. The x-axis depicts V/J, which is the similarly normalized quadrupolar interaction 
defined as 

\begin{equation}
V = (9/2) Q^2/a^5 
\end{equation}

\noindent
where $Q$ is the quadrupole moment and $a$ is the unit cell constant. This is 
a repulsive term involving $d$ orbitals on nearest-neighbor sites and $Q$ increases as a function of the degree of “hybridization” 
of oxygen $p$- and metal $d$-orbitals.  Of the seven states, three are ferromagnetic: FM110, FM111 and *. 
There are also three antiferromagnetic states: AFM100, $\Delta$, and $\overline{\Delta}$; the latter two represent
 complex four-sublattice structures.  The final state is a quadrupolar state, also described as a “spin nematic”, 
which does not break time reversal symmetry. 

 \begin{figure}[ht]
 \includegraphics[width=85mm]{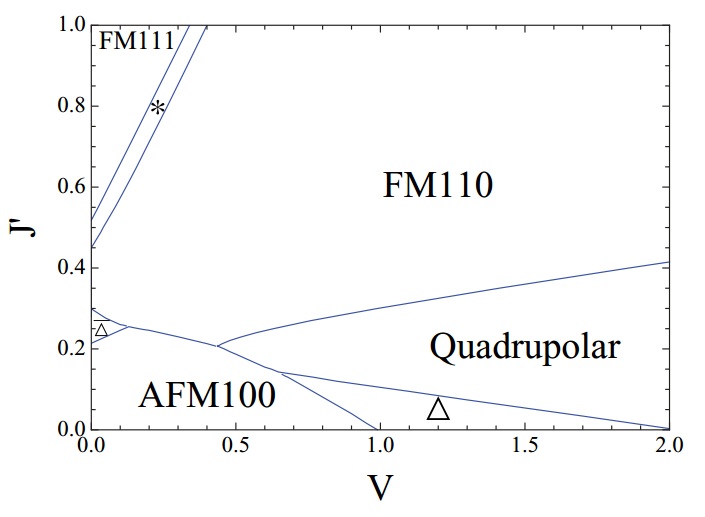}
 \caption{\label{}Calculated phase diagram of $d^2$ double perovskites exhibiting spin-orbit coupling (SOC), under
the model Hamiltonian H = H$_{AFM}$ + H$_{FM}$ + H$_{quad}$.   
The axes represent J', a nearest-neighbor FM interaction, and V, the quadrupolar interaction defined in the text. 
Both are normalized by J, representing the nearest-neighbor AFM interaction.   Reproduced from Figure 1 in Ref. \cite{balents_s_1}; 
copyright (2011) by the American Physical Society.}
 \end{figure}

 It is of interest of course to determine the relative placement of Ba$_2$CaOsO$_6$ and Ba$_2$YReO$_6$ 
within this diagram, $i.e.$, with respect to the J'/J and V/J axes. Comparing the Curie-Weiss temperatures, both are 
strongly negative but $\Theta$$_\mathrm{(Ba_2YReO_6)}$/ $\Theta$$_\mathrm{(Ba_2CaOsO_6)}$ $\sim$ 4, indicating 
that J'/J$_{\mathrm{(Ba_2YReO_6)}}$ $<$  J'/J$_{\mathrm{(Ba_2CaOsO_6)}}$. To determine the relative V/J positions, 
the Re$-$O and Os$-$O distances should reflect the degree of hybridization 
with oxygen. These are 1.955(1) \AA\ and 1.919(2) \AA,
 respectively, indicating a larger V/J for the Ba$_2$CaOsO$_6$ phase.  With the present work and previous heat capacity 
studies \cite{yamamura} the ground state of the CaOs material is well established as AFM and limits can be 
placed on the magnitude of the ordered Os$^{6+}$ moment using both neutron diffraction and $\mu$SR results. 
Unfortunately, the type of AFM order cannot yet be determined due to the very small ordered moment. 
 Ba$_2$CaOsO$_6$ is thus  described by the small J'/J part of Figure 11, and the likely choices are AFM100, $\Delta$ or 
$\overline{\Delta}$, the latter two being complex four-sublattice states.

There is one further important result from Ref. \cite{balents_s_1} which can guide this choice.  
The calculations were also carried out for $T$ $>$ 0 and thus include
 predictions concerning the nature of the phase transitions from the high temperature paramagnetic state 
to any of the $T$ = 0 phases of Fig. 11. The transitions to $\Delta$ and $\overline{\Delta}$ are found to 
be continuous while that to AFM100 is strongly 
first order. The $\mu$SR data of Fig. 8 indicate a continuous transition for Ba$_2$CaOsO$_6$ and thus there is a strong
argument for either the $\Delta$ or $\overline{\Delta}$ ground states.   Indeed, it was suggested that Ba$_2$CaOsO$_6$
might be assigned to $\overline{\Delta}$.   In the same paper it was postulated that Ba$_2$YReO$_6$ could be a candidate
for the quadrupolar-spin nematic region of the phase diagram.   However, this seems inconsistent, if
V(Ba$_2$CaOsO$_6$) $>$ V(Ba$_2$YReO$_6$), as argued here.   For any of the above choices of ground state for Ba$_2$CaOsO$_6$,
given that both V/J and J'/J are smaller for Ba$_2$YReO$_6$, the AFM100 state would be expected for this material,
according to Fig. 11.   It was suggested that disorder precludes the establishment of one of the predicted ground states for
Ba$_2$YReO$_6$.  However, the level of one type of potential disorder, namely Y/Re intersite exchange, was examined using 
the very sensitive technique of $^{89}$Y magic angle spinning (MAS) NMR and such disorder was not detected, 
placing an upper limit of $\lesssim$0.5 \% on B/B' site disorder 
\cite{tomoko_1}.  Interestingly, Y/M site disorder was easily seen by this method in the closely related double perovskites, Ba$_2$YMO$_6$ 
 at the levels of 1\% (M = Ru) and 3\% (M= Mo) \cite{tomoko_3_2, tomoko_1_2}. 
The ground states of these two materials appear to be unaffected by such larger disorder levels, 
as seen in Table I.    That an ordered ground state is not found for Ba$_2$YReO$_6$ indicates either that a very low, as yet
undetectable, disorder level plays a decisive role or that one must go beyond the mean theory level of Ref. \cite{balents_s_1}
to find an explanation.

\section{Conclusions}

The double perovskite Ba$_2$CaOsO$_6$, based on the 5$d^2$ ion, Os$^{6+}$, has been synthesized and 
characterized using neutron diffraction and $\mu$SR techniques which complement earlier studies of the heat 
capacity and magnetic susceptibility \cite{yamamura}.  It is now clear that cubic $Fm\overline{3}m$ symmetry is retained to 
3.5~K and that the material orders antiferromagnetically near 50~K with an estimated ordered moment size 
of $\sim$0.2$\mu_\mathrm{B}$. However, the long-range nature of the ordered state cannot be determined solely 
from the $\mu$SR data, as $\mu$SR is a local probe sensitive only to local field environments.  This 
behaviour is in remarkable contrast to that of its  isostructural and isoelectronic d\"{o}ppelganger, Ba$_2$YReO$_6$, 
which, with essentially the same cubic unit cell constant, has a $\Theta_{CW}$ four times larger and is 
found in a spin frozen ground state with $T_g$~$\sim$~35~K \cite{tomoko_1}.  The behaviour of Ba$_2$CaOsO$_6$ 
can also be contrasted with the closely related 4$d^3$ and 4$d^1$ based double perovskites, Ba$_2$YRuO$_6$ 
and Ba$_2$YMoO$_6$, which also exhibit cubic $Fm\overline{3}m$ symmetry and have AFM \cite{tomoko_3_2,carlo_BYRO} 
and gapped spin singlet ground states \cite{deVries_BYMO,tomoko_1_2,carlo_BYMO} respectively.  This is consistent
of a general evolution from ordered to increasingly disordered ground states as one proceeds from $d^3$ to $d^2$ 
to $d^1$ systems.   As monoclinic La$_2$LiReO$_6$ also does not order \cite{tomoko_1}, Ba$_2$CaOsO$_6$ is the 
first 5$d^2$ based double perovskite which does, exhibiting $\mu$SR spectra more closely resembling the 4$d^3$ 
Ba$_2$YRuO$_6$ than the isoelectronic 5$d^2$ system Ba$_2$YReO$_6$.
    
The properties of Ba$_2$CaOsO$_6$ and Ba$_2$YReO$_6$ have been discussed in the context of a recently 
published mean field theory for n$d^2$ double perovskites under the influence of  strong SOC \cite{balents_s_1}.  
While Ba$_2$CaOsO$_6$ might be described in terms of an exotic four sublattice AFM state, $\Delta$ or $\overline{\Delta}$,
it is difficult to accommodate Ba$_2$YReO$_6$ within this theoretical framework.  While some level of disorder might play a role in the 
latter case, there is scant experimental evidence.

\ack

J.E.G. thanks the Natural Sciences and Engineering Research Council of Canada (NSERC) for support through the 
Discovery Grant program, and Dr. E. Kermarrec for useful discussions.  G. M. L. acknowledges support from NSERC.
The Columbia group has been supported by US NSF 
under DMR-1105961 and OISE-0968226 (PIRE: Partnership for International Research and Education) and by 
JAEA (Japan) under the REIMEI project.   S. C. acknowledges travel support from the Friends of Todai Foundation.
We thank the TRIUMF CMMS staff for invaluable technical assistance with $\mu$SR experiments.

\section*{References}


\end{document}